\documentstyle[12pt]{article}
\def\journal#1, #2, #3, #4 { {\sl #1~}{\bf #2~} #3 #4 }

\def\cmp{\journal Comm. Math. Phys., }

\def\np{\journal Nucl. Phys., }

\def\pl{\journal Phys. Lett., }

\def\ijmp{\journal Int. J. Mod. Phys., }
\def\beq{\begin{equation}}
\def\eeq{\end{equation}}
\def\beqa{\begin{eqnarray}}
\def\eeqa{\end{eqnarray}}
 \def\nnn{\nonumber \\}
\def\sqr#1#2{{\vcenter{\vbox{\hrule height.#2pt
\hbox{\vrule width.#2pt height#1pt \kern#1pt
\vrule width.#2pt}
\hrule height.#2pt}}}}

\def\Jhat{{\widehat J}}

\def\mhat{{\widehat m}}

\def\mhat{{\widehat m}}

\begin{document}
\begin{titlepage}

\nopagebreak
\begin{flushright}

LPTENS--93/19\\
hep-th@xxx/9305043 \\
    May 1993
\end{flushright}

\vglue 1  true cm
\begin{center}
{\large \bf
THE BRAIDING OF CHIRAL VERTEX OPERATORS \\
\medskip
WITH CONTINUOUS  SPINS  IN 2D GRAVITY }
\vglue 1.5 true cm
{\bf Jean-Loup~GERVAIS}\\
\medskip
and \\
\medskip
{\bf Jens SCHNITTGER}{\footnotesize\footnote{supported by DFG}}\\
\medskip
{\footnotesize Laboratoire de Physique Th\'eorique de
l'\'Ecole Normale Sup\'erieure\footnote{Unit\'e Propre du
Centre National de la Recherche Scientifique,
associ\'ee \`a l'\'Ecole Normale Sup\'erieure et \`a
l'Universit\'e
de Paris-Sud.},\\
24 rue Lhomond, 75231 Paris CEDEX 05, ~France}.
\end{center}
\vfill
\begin{abstract}
\baselineskip .4 true cm
\noindent
{\footnotesize Chiral vertex-operators are defined for continuous
quantum-group spins $J$ from  free-field realizations  of the Coulomb-gas
type. It is shown that these generalized chiral vertex operators
satisfy closed braiding relations on the unit circle, which are
given by an extension in terms of orthogonal polynomials
of the braiding matrix recently derived by
Cremmer, Gervais and Roussel.
This leads to a
natural extension of the
Liouville exponentials to continuous powers that remain local.
}
\end{abstract}
\vglue 3 true cm
\vfill
\end{titlepage}

\section {INTRODUCTION}
Recently there has been considerable progress in  understanding
the structure of two-dimensional gravity from the continuum point
of view\cite{B}-\cite{CGR}.
It was brought about by the realization of the underlying
quantum group structure of the theory which fully determines the chiral
operator algebra\cite{B} \cite{G1} \cite{G3} \cite{CGR}.
The basic chiral fields from which the local
Liouville operators $\exp (-\alpha J\Phi)$
are constructed ($\alpha$ is one of the screening charges)
 were seen to fall into  standard
representations\footnote{to simplify, we
only deal with irrational theories
explicitly,   so
that these  representations are in a one-to-one correspondence with those
of $sl(2)$} of $U_q(sl(2))$ for  (positive) half-integer spin $J$.
They yield a
comprehensive description of the "minimal" sector of the theory consisting
only of degenerate primaries. Correspondingly, the derivation
of refs.\cite{G1} \cite{CGR}  was
largely based on the null-vector decoupling-equations  for the simplest
degenerate primaries, which arise naturally upon quantization of
the Liouville dynamics. Further progress was made\cite{G2} \cite{G3}
\cite{G5}  by  putting forward
 the symmetry $J\rightarrow -J-1$ between positive and negative spins
which already leads
 out of the "minimal" sector and allows to construct
the three-point function of minimal matter coupled to gravity\cite{G5},
 as well as
to derive unitary-truncation theorems in strong coupling
theories\cite{G3} \cite{GR1} \cite{GR2}. However,
for a full understanding of the integrability structure of the theory,
and, in particular,  for
the construction of arbitrary Liouville exponentials
$e^{\lambda  \varphi}$ (and thus the Liouville field itself as
$\varphi = {d\over d\lambda}|_{\lambda =0} e^{\lambda \varphi}$)
 it is  necessary to consider
the general case of continuous  spins. This will provide information
about the structure of Liouville theory as a local conformal theory
in its own right --- not just as a "gravitational dressing" for some
particular matter theory --- which seems hardly accessible from the
matrix model approach. Such insights  appear particularly useful
to understand 2 dimensional gravity, and its W generalizations
in their  strong-coupling regimes  where, so far, only
the quantum-group approach has led to significant progress\cite{G2}
\cite{G3} \cite{GR1} \cite{GR2}.
 In this paper
we shall take a first  step in this direction by establishing
control over the exchange algebra of chiral primaries with
continuous  spins, which is the essential ingredient for the construction
of general Liouville exponentials. In the absence of an underlying
null-vector decoupling equation, we shall rely exclusively
on free field techniques, pushing further previous
attempts\cite{OW}
which were confined to the lowest orders in the cosmological constant
and did not disentangle the chiral algebra which is ultimately
responsible for the locality properties of the Liouville field.
The emerging picture will  be in accord with the naive expectation
that,   in the general situation, the algebra of the chiral primaries
should be governed by the universal $R$-matrix of $SL(2)_q$,
suitably generalized  to
(infinite-dimensional) highest/lowest-weight representations with
continous spins.
 Our notations are identical to those of refs.\cite{G1}-\cite{CGR}.
The chiral primaries $V_m^{(J)}$ in the Bloch-wave basis are the
fields with diagonal  monodromy ---
 as opposed to the fields $\xi_M^{(J)}$
of the quantum-group basis, which transform  as multiplets of
spin $J$ under quantum-group action\footnote{In the language of
integrable lattice models, the former lead to
interaction-around-the-face models, while the latter describe
interactions of the vertex type.}.
 In this letter, we  make use of
one of the two screening charges only\footnote{of course, this
does not matter since we will eventually take $J$ to be
continuous.},
so that we start from
$V_m^{(J)}$, and
not from the most general fields $V_{m \mhat} ^{(J \Jhat)}$.
In ref.\cite{G5}, it was shown that, for $2J$ a positive integer,
the exponential of the Liouville field is given by an expression of
the form
\begin{equation}
e^{\textstyle -J\alpha_-\Phi(\sigma, \tau )}=
\>\sum _{m=-J}^J \,
B_m^{(J)}(\varpi)\>
V_m^{(J)}(x_+)\,
{\overline V_{m}^{(J)}}(x_-)
\label{1.1}
\end{equation}
where $\sigma$, and $\tau$ are world-sheet variables,
$x_\pm=\sigma\mp i\tau$, $\varpi$ is the Liouville zero
mode\footnote{appropriately rescaled}, and ${\overline V_{m}^{(J)}}$
are  the antiholomorphic components. The function $B_m^{(J)}(\varpi)$
may be deduced from the expression given in ref.\cite{G5} ---
it is not explicitly
needed at this point. It is determined from the requirement that
two Liouville
exponentials commute at different $\sigma$, for equal $\tau$, as
required by locality.   This neatly follows from the braiding
relations of the $V$ fields which are given by\cite{CGR}:
\beq
V_m^{(J)}(\sigma) \> V_{m'}^{(J')}(\sigma')=
\sum_{m_1 m_2} \hat {R}(J,J';\varpi)_{m_{\phantom{2}} m'}^{m_2 m_1}
V_{m_2}^{(J')}(\sigma') \> V_{m_1}^{(J)}(\sigma)
\label{1.2}
\eeq
\beq
\hat {R}(J,J';\varpi)_{m_{\phantom{2}} m'}^{m_2 m_1} =
q^{2mm' + m'^2 -m_2^2 + \varpi (m'-m_2) }
{g^{x+m_2}_{J,x+M} g^x_{J',x+m_2} \over g^{x+m}_{J',x+M} g^x_{J,x+m}}
\left\{ ^{J}_{J'}\,^{x+M}_{x}
\right. \left |^{x+m_2}_{x+m}\right\}
\label{1.3}
\eeq
$$
M=m+m'=m_1+m_2, \quad
x:=(\varpi -\varpi_0)/2, \quad  \varpi_0 :=1+\pi/ h, \quad q:=e^{ih},
$$
\beq
h={\pi \over 12}\Bigl(C-13 -
\sqrt {(C-25)(C-1)}\Bigr),
\label{1.4}
\eeq
where $\left\{ ^{J}_{J'}\,^{x+M}_{x} \right.
 \left |^{x+m_2}_{x+m}\right\}$ denotes  the quantum 6-j symbol,
$C$ is the central charge, and
the $g$'s are coupling constants whose
 expression  will be
recalled below when needed\footnote{For an earlier derivation in a different
form, see also \cite{FFK}}.
In the approach of ref.\cite{G5}, one takes
the zero modes of the left-moving and right-moving Liouville modes
to commute, so that $V$ and $\bar V$ commute. Thus  the equations just
written, together with their right-moving counterparts
are the first step  to
ensure the locality properties. Since one considers the braiding at
equal $\tau$ one lets $\tau=0$ once and for all, and works  on the unit
circle $u=e^{i\sigma}$. In the above equations
and in the following we discuss the braiding of the
holomorphic components assuming that    $\sigma$ and $\sigma'$
range between $0$ and $\pi$ for concreteness. Our task in the
present letter is to generalize Eqs.\ref{1.2} and \ref{1.3} to
continuous $J$.
To construct the chiral primaries, one introduces two equivalent
free fields\cite{GN3}
\beq
\phi_j(\sigma)=q^{(j)}_0+ p^{(j)}_0\sigma+i
\sum_{n\not= 0}e^{-in\sigma}\, p_n^{(j)}\bigl / n,
\quad j=1,\> 2,
\label{1.5}
\eeq
such that (primes mean derivatives) \hfill
\[
\Bigl [\phi'_1(\sigma_1),\phi'_1(\sigma_2) \Bigr ]=
\Bigl[\phi'_2(\sigma_1),\phi'_2(\sigma_2) \Bigr ]
=2\pi i\,  \delta'(\sigma_1-\sigma_2),
\]
\beq
p_0^{(1)}=-p_0^{(2)}=-i\sqrt{h/2\pi}\varpi ,
\label{1.6}
\eeq
Chiral primaries $V_m^{(J)}$ with
arbitrary $m$ and $J$ can then be defined as fusions of exponentials
of $\phi_1$ and $\phi_2$:
\beq
V_m^{(J)} = K_m^{(J)} (\varpi)
N^{(1)}\bigl (e^{\sqrt{h/2\pi}(J-m) \>\phi_1}\bigr )\bullet
N^{(2)}\bigl (e^{\sqrt{h/2\pi}(J+m) \>\phi_2}\bigr ),
\label{fusion}
\eeq
Here $N^{(j)}$ denotes normal
ordering
with respect to the modes of $\phi_j$, and $\bullet $ abbreviates
the renormalized short-distance product (fusion)
of the two factors\cite{BiG}. $K_m^{(J)}$ is a normalization factor such that
the ground state matrix elements of $V_m^{(J)}$ are canonically normalized:
\beq
\langle \varpi |V_m^{(J)}(\tau =\sigma =0) |\varpi +2m \rangle =1
\label{norm}
\eeq
In particular, one may see
that $V_{\pm J}^{(J)}$ are  expressible as simple exponentials:
 \beq
V_{- J}^{(J)} = N^{(1)}\bigl (e^{\sqrt{h /2\pi}\>
2J \phi_1}\bigr
), \>
 \quad
V_{ J}^{(J)} = N^{(2)}\bigl (e^{\sqrt{h /2\pi}\>
2J \phi_2}\bigr
).
\label{1.8}
\eeq
They are simple ``tachyon'' vertex operators. One may of course
verify that, for $J\pm m=0$, and $J'\pm m'=0$, Eq.\ref{1.2} does
reduce to the simple braiding of exponentials  of free fields.

\section{Derivation of the chiral algebra for continous~J}

For the reconstruction of the Liouville exponentials with
arbitrary $J$ generalizing Eq.\ref{1.1}, one needs only the chiral
primaries with $J+m$  (positive)
integer, or equivalently, those with $J-m$ integer. This
can be seen already from the classical formula\cite{G4} and will
be shown in detail in
a later publication.  We shall only discuss
the case of integer $J+ m$  explicitly here. The
other case  may  be obtained by the trivial substitution $ \phi_1 \to
\phi_2  $ and
$\varpi \rightarrow -\varpi$ in all the formulae. In this situation,
the Bloch-wave operator $V_{m}^{(J)}$
may be re-expressed in terms of $\phi_1$ alone by means of a
Coulomb-gas type integral representation. It has the form\cite{LS}
\beq
V_{m}^{(J)} =(I_{m}^{(J)}(\varpi) )^{-1} U_{m}^{(J)},
\label{2.0}
\eeq
where
$I_{m}^{(J)}(\varpi)$ are normalization constants (independent of
$\sigma$), and
\beq
U_{m}^{(J)}(\sigma)=  V_{-J}^{(J)}(\sigma) [S(\sigma)]^{J+m}
\label{2.1}
\eeq
\beq
S(\sigma)= e^{2ih(\varpi+1)}
 \int _0 ^{\sigma} dx V_{1}^{(-1)}(x)+
 \int _\sigma  ^{2\pi } dx V_{1}^{(-1)}(x).
\label{2.2}
\eeq
Indeed, first  it is easily seen that $V_{1}^{(-1)}$ is one of the two
screening operators. Second,  using the explicit expression
for the Liouville zero-mode $\varpi=ip_0^{(1)}\sqrt {2\pi/h}$
(see Eq.\ref{1.6}),
one verifies that
\beq
V_m^{(J)} \varpi = (\varpi+2m) V_m^{(J)}
\label{2.3}
\eeq
as expected.  Third the  formula  for $S(\sigma)$ is such that
\beq
S(\sigma+2\pi)=e^{2ih(1+\varpi)}S(\sigma).
\label{2.4}
\eeq
 From this, one deduces that
\beq
U_{m}^{(J)}(\sigma+2\pi)= e^{2ihm\varpi+2ih m^2}
U_{m}^{(J)}(\sigma)
\label{2.5}
\eeq
which gives   the correct monodromy eigenvalue
  for the fields $V_m^{(J)}$.
Since the above
expressions make sense also for continuous $2J$, provided that
$J+m$ is a non-negative integer,
we take them as the definition of the generalized
vertex operators. The only potential problem comes from
divergences of the product of $S(\sigma)$. However it is easily
seen that $[S]^{J+m}$ makes sense for small enough $h$. The singularities
which appear when $h$ increases are to be handled by analytic
continuation (more about this below).
We now turn to the derivation of the algebra of the fields
$U_m^{(J)}$.
The quantum group picture leads us to expect that there should in fact
exist a closed exchange algebra for these operators which is related
to the braiding of highest (lowest) weight representations of
$U_q(sl(2))$
with continous spin. Hence we start with an ansatz of the form
\beqa
U_m^{(J)}(\sigma) U_m^{(J')}(\sigma')=&
\sum_{m_1,m_2} R(J,J';\varpi)^{m_2 m_1}_{m\phantom{_2} m'\phantom{_1}}
U_{m_2}^{(J')}(\sigma') U_{m_1}^{(J)}(\sigma)
\nnn
&\hbox{for}\quad \pi >\sigma '>\sigma >0,
\nnn \hbox{resp.}&
\nnn
U_m^{(J)}(\sigma) U_m^{(J')}(\sigma')=&
\sum_{m_1,m_2} \tilde R(J,J';\varpi)^{m_2 m_1}_{m\phantom{_2} m'\phantom{_1}}
U_{m_2}^{(J')}(\sigma') U_{m_1}^{(J)}(\sigma)
\nnn
&\hbox{for} \quad \pi >\sigma >\sigma  ' >0,
 \label{2.6}
\eeqa
with the sums extending over non-negative integer $J+m_1$ resp. $J'+m_2$.
Comparing the monodromy properties of both sides of Eq.\ref{2.6}, we conclude
that,  as in the half-integer spin case, the braiding matrix is nonzero
only when
\beq
m_1+m_2=m+m'=:M
\label{2.7}
\eeq
Since there are no  null-vector decoupling equations for continuous $J$,
 we have to rely exclusively on the free
field techniques  just summarized.
 Fortunately, it turns out that the exchange of two $U_m^{(J)}$
operators can be mapped into an equivalent problem in one-dimensional
quantum mechanics, and becomes just finite-dimensional linear algebra.
Given the fact that the $U_m^{(J)}$ consist only of
the "tachyon operators" $V_{-J}^{(J)}$
resp. $V_{1}^{(-1)}$, the essential observation
is to remember that
\beq
V_{-J}^{(J)}(\sigma )V_{-J'}^{(J')}(\sigma ')=
e^{-i2JJ' h \epsilon (\sigma -\sigma')}
V_{-J'}^{(J')}(\sigma ' )V_{-J}^{(J)}(\sigma )
\label{2.8}
\eeq
where $\epsilon (\sigma -\sigma')$ is the sign of $\sigma -\sigma'$.
This means that when commuting the tachyon operators in
$U_{m'}^{(J')}(\sigma ')$ through those of
$U_{m}^{(J)}(\sigma )$, one only encounters phase factors of the form
$e^{\pm i2 \alpha \beta h } $ resp. $e^{\pm  6i \alpha \beta h } $,
with $\alpha$ equal to $J$ or $-1$, $\beta$ equal to $J'$ or $-1$,
since  we take
$\sigma, \sigma' \in  [0,\pi]$. Hence we are led to decompose the
integrals defining the screening charges S into pieces which commute
with each other and  with $V_{-J}^{(J)}(\sigma)$, $V_{-J'}^{(J')}
(\sigma ')$ up to one of the above phase factors. We consider
explicitly only the case $0< \sigma < \sigma ' < \pi$ and write
\[
S(\sigma )\phantom{'} = S_{\sigma \sigma '} + S_\Delta,
\quad
S(\sigma ')= S_{\sigma \sigma '} + k(\varpi)S_\Delta \equiv
S_{\sigma \sigma '} + \hat {S}_\Delta,
\]
\[
S_{\sigma \sigma '}:= k(\varpi)
\int_0^\sigma V_{1}^{(-1)}(\tilde \sigma )
d\tilde \sigma + \int_{\sigma '}^{2\pi}V_{1}^{(-1)}(\tilde \sigma )
d\tilde \sigma,
\]
\beq
S_\Delta := \int_\sigma^{\sigma '}V_{1}^{(-1)}(\tilde \sigma )
d\tilde \sigma,
\quad
k(\varpi):= e^{2ih(\varpi+1)}
\label{2.9}
\eeq
Using Eq.\ref{2.8}, we then get the following simple algebra for
$S_{\sigma \sigma '}, S_\Delta , \hat {S}_\Delta$:
\medskip
\beq
S_{\sigma \sigma '}S_\Delta =q^{-2}S_\Delta S_{\sigma \sigma '},
\quad
S_{\sigma \sigma '}\hat {S}_\Delta =q^2 \hat {S}_\Delta
S_{\sigma \sigma '},
\quad
S_\Delta \hat {S}_\Delta =q^4 \hat {S}_\Delta S_\Delta,
\label{2.10}
\eeq
and their commutation properties with
$V_{-J}^{(J)}(\sigma ),V_{-J'}^{(J')}(\sigma ' )$ are given by
 \beq
\begin{array}{lll}
&V_{-J}^{(J)}(\sigma) S_{\sigma \sigma '}=q^{-2J}S_{\sigma \sigma'}
V_{-J}^{(J)}(\sigma ),
&V_{-J'}^{(J')}(\sigma')S_{\sigma \sigma '}= q^{-2J'} S_{\sigma \sigma'}
V_{-J'}^{(J')}(\sigma'),
\nnn
&V_{-J}^{(J)}(\sigma)S_\Delta  = q^{-2J}S_\Delta V_{-J}^{(J)}(\sigma ),
&V_{-J}^{(J)}(\sigma) \hat {S}_\Delta \  =q^{-6J} \hat {S}_\Delta
V_{-J}^{(J)}(\sigma),
\nnn
&V_{-J'}^{(J')}(\sigma')S_\Delta = q^{2J'} S_\Delta V_{-J'}^{(J')}
(\sigma '),
&V_{-J'}^{(J')}(\sigma')\hat {S}_\Delta = q^{-2J'} \hat {S}_\Delta
V_{-J'}^{(J')}(\sigma').
\end{array}
\label{2.11}
\eeq
Finally, all three screening pieces obviously shift the zero mode in the
same way:
\beq
 \left. \begin{array}{ccc}
S_{\sigma \sigma '}\nnn S_\Delta \nnn  \hat {S}_\Delta
\end{array} \right \} \varpi
= (\varpi +2) \left \{ \begin{array}{ccc}
S_{\sigma \sigma '}\nnn  S_\Delta \nnn  \hat {S}_\Delta
\end{array} \right. .
\label{2.12}
\eeq
Using Eqs.\ref{2.11} we can commute $V_{-J}^{(J)}(\sigma  )$ and
$V_{-J'}^{(J')}(\sigma  ')$ to the left on both sides of  Eq.\ref{2.6},
so that they can be cancelled. Then we are left with

\[
(q^{-2J'}S_\Delta + q^{2J'}S_{\sigma \sigma '} )^{J+m}
(\hat {S}_\Delta +S_{\sigma \sigma '})^{J'+m'}=
\]
\beq
\sum_{m_1,m_2} R(J,J';\varpi + 2(J+J'))_{m_{\phantom{2}}m'}^{m_2 m_1}
q^{-2JJ'}
(q^{2J} S_{\sigma \sigma '} + q^{6J}\hat {S}_\Delta )^{J'+m_2}
(S_{\sigma \sigma '} + S_\Delta )^{J+m_1}
\label{2.13}
\eeq
It is apparent from Eq.\ref{2.13} that the braiding problem of the
$U_m^{(J)}$ operators is governed by the Heisenberg-like algebra
Eq.\ref{2.10}, characteristic of one-dimensional
quantum mechanics. However,
to see this structure emerge, we had to decompose the screening charges
$S(\sigma ), S(\sigma ')$ in a way which depends on both positions
$\sigma ,\sigma '$; hence the embedding of this Heisenberg algebra
into the 1+1 dimensional field theory is somewhat nontrivial.
To evaluate Eq.\ref{2.13}, we could sort both sides of the
equation in powers
of $S_{\sigma \sigma '},S_\Delta $ and then compare coefficients. This
indeed can be carried out straightforwardly, upon observing that
$S_{\sigma \sigma '}$ and $S_\Delta$ resp. $\hat {S}_\Delta $ behave
like the components $a,b$ of a vector in the quantum plane, with
$ba=abq^2$, so that one can make use of the q-binomial formula
\[
(a+b)^N = \sum_{\nu =0}^N \hbox {${N \choose \nu}$} q^{(N-\nu )\nu}
a^\nu b^{N-\nu},\quad
{N \choose \nu}:= { \lfloor N \rfloor ! \over \lfloor N-\nu
\rfloor ! \lfloor \nu \rfloor !}
\]
${N \choose \nu}$ is a q-deformed binomial coefficient, with
\beq
\lfloor \nu \rfloor != \lfloor 1 \rfloor \lfloor 2 \rfloor  \cdots
\lfloor \nu \rfloor, \quad \lfloor x \rfloor :=\sin(hx) / \sin(h)
\label{2.14}
\eeq
denoting q-factorials resp. q-numbers.
For our purposes, however, another form of the equations is better
suited, which is obtained by choosing the following simple representation
of the algebra Eq.\ref{2.10} in terms of one-dimensional
quantum mechanics ( $y$ and $y'$ are arbitrary complex numbers):
\beq
S_{\sigma \sigma '}=y' e^{2Q}, \quad
S_\Delta =y e^{2Q-P}, \quad  \hat {S}_\Delta =
y e^{2Q+P},
 \qquad [Q,P]=ih  .
\label{2.15}
\eeq
The third  relation in Eq.\ref{2.15} follows from
the second one in view of
$\hat {S}_\Delta =k(\varpi )S_\Delta$ (cf. Eq.\ref{2.9}).
	This means we are
identifying here $P \equiv ih\varpi $ with the zero mode of the original
problem. Using $e^{2Q+cP} = e^{cP} e^{2Q} q^c$ we can commute all factors
$e^{2Q}$ to the right on both sides of Eq.\ref{2.13} and then cancel them.
This leaves us with
\[
\prod_{j=1}^{J+m}(y'q^{2J'} +yq^{-(\varpi -2J+2j-1)})
\prod_{\ell=1}^{J'+m'} (y'+yq^{\varpi -2J'+2m+2\ell-1})=
\]
\[
\sum_{m_1} R(J,J';\varpi)_{m_{\phantom {2}} m'}^{m_2 m_1} q^{-2JJ'}
\times
\]
\beq
\prod_{j=1}^{J'+m_2}(y'q^{2J} +yq^{\varpi +4J-2J'+2j-1})
\prod_{\ell=1}^{J+m_1} (y'+yq^{-(\varpi -2J+2m_2 +2\ell -1)})
\label{2.16}
\eeq
where we have shifted back $\varpi +2(J+J') \rightarrow \varpi$
compared to Eq.\ref{2.13}.
Since the overall scaling $ y\rightarrow \lambda y,
y' \rightarrow \lambda y'$ only gives back Eq.\ref{2.7}, we can set $y'=1$.
By putting $y$ equal to the zeros of the first or the second product
on the RHS of Eq.\ref{2.16}, plus one other arbitrary value, we obtain a
linear system of equations for $R$ in triangular form, which shows that
the solution of Eq.\ref{2.16} is unique for any fixed $J,J',m,m'$.
We will now demonstrate that it is
given by a straightforward extension of the
R-matrix of ref.\cite{CGR} (see Eq.\ref{1.3}).
This extension can be written in terms of orthogonal
polynomials and we prove its correctness by showing that Eq.\ref{2.16} is
equivalent to the orthogonality relations for these polynomials.
According to the beginning of this section, the braiding matrices of
the $V$ and $U$ fields are related by
\beq
R(J,J';\varpi)_{m_{\phantom{2}}m'}^{m_2 m_1} =\hat {R}(J,J';\varpi )
_{m_{\phantom{2}}m'}^{m_2 m_1} {I_m^{(J)}(\varpi ) I_{m'}^{(J')}(\varpi
 +2m) \over I_{m_2}^{(J')}(\varpi) I_{m_1}^{(J)}(\varpi +2m_2) }
\label{2.17}
\eeq
The normalizations $I_m^{(J)}$ can be computed by making use of the
 Fateev-Dotsenko integration formulae\cite{DF}.
One finds\footnote{details will be given elsewhere.}, letting
$n=J+m$,
\[
I_m^{(J)}(\varpi )=i^n \prod_{l=1}^n \left \{e^{i\pi  \beta (l-1)}
(1-e^{2\pi i (\gamma +  \beta (l-1))})\right \} \times  $$
$$\prod_{l=1}^n \left \{{\Gamma (1-\beta ) \Gamma (1+\gamma +(l-1)\beta )
\Gamma (1+\alpha +(l-1)\beta )
\over \Gamma (1-l\beta ) \Gamma (2+\gamma +\alpha +(n-2+l)\beta )}\right \}
\]
\beq
\alpha = 2J{h \over \pi}, \quad
\beta = -{h \over \pi}, \quad
\gamma={h \over \pi} (\varpi +2m -1) -1.
\label{2.18}
\eeq

This relation  is valid for arbitrary $J,\varpi $ and $J+m=n$ a
non-negative integer.
We remark that the divergence of the product
$[S]^{J+m}$ appears as the first pole in $h$ in the formula just written.
It shows  where
the integral representation Eq.\ref{2.9} breaks down. However,
Eq.\ref{2.18} has meaning beyond this point by the usual
analytic continuation
of the gamma function. Indeed, the general formula Eq.\ref{fusion} is valid
for arbitrary $h$ and will give rise to a 3-point function which is
analytic in $h$. Hence Eq.\ref{2.18} is valid for all $h$.
Next we consider the
formula for the coupling constant
$g^{J_{12}}_{J_1 J_2}$  of ref.\cite{CGR}.
It is given by (letting $F(z)\equiv \Gamma(z)/\Gamma(1-z)$)
$$
g_{J_1J_2}^{J_{12}}
=
\prod_{k=1}^{J_1+J_2-J_{12}}\sqrt{ F(1+(2J_1-k+1)h/\pi)}\times
$$
\beq
\sqrt{
F(1+(2J_2-k+1)h/\pi)
F(-1-(2J_{12}+k+1)h/\pi)
\over
F(1+kh/\pi)
}
\label{2.19}
\eeq
It is easily seen that for the general
$J$  case,    $J_1 + J_2 - J_{12} $ remains an integer, so that
this expression directly  makes sense.
Thus we only
 need to specify  the proper continuation
of the 6j-symbol. Using the relation between the 6j-symbol and the
$_4 F_3$ q-hypergeometric function\cite{KR}\cite{HHM}\cite{CGR},
one can write\footnote{our conventions for
q-hypergeometric functions  coincide with the ones of
ref.\cite{G3}.}
$$
\left\{ ^{J}_{J'}\,^{x+M}_{x}
\right. \left |^{x+m_2}_{x+m}\right\}=\sqrt{\lfloor 2x+2m_2 +1\rfloor
\lfloor 2x + 2m +1\rfloor }\times
$$
$$
\Delta (J,x+M,x+m_2)\Delta (J,x,x+m) \Delta (J',x,x+m_2) \Delta (J',x+m,x+M)
\times
$$
$$
{\lfloor 2x+N+1\rfloor ! \over \lfloor 2x + m_2 +m-J-J'\rfloor !
\lfloor J+m_1 \rfloor !\lfloor J-m\rfloor !\lfloor J'+m'\rfloor ! \lfloor
J'-m_2 \rfloor ! }
\times
$$
\beq
{1\over \lfloor m_2-m'\rfloor !} \> _4 F_3 \left (^{m-J,}_{-(2x+N+1),} \,
^{m_2-J',}_{m_2-m'+1,} \, ^{-(J+m_1),}_
{2x+m_2 +m+1-J-J'} \, ^{-(J'+m')} \ ;q,1 \right )
\label{2.20}
\eeq
with  \hfill
\[
N=J+J'+M,   \qquad
\Delta (a,b,c) =\sqrt{{\lfloor -a+b+c\rfloor !\lfloor a-b+c\rfloor !
\lfloor a+b-c\rfloor !\over \lfloor a+b+c+1 \rfloor !}}
\]
The RHS of Eq.\ref{2.20} makes sense for arbitrary $J$. The
q-hypergeometric function is defined as
\[
_4 F_3 \left (^{a,}_{e,} \, ^{b,}_{f,} \,
^{c,}_{g} \, ^{d} ;q,\rho \right )=
\sum_{n=0}^\infty {\lfloor a\rfloor_n \lfloor b\rfloor_n
\lfloor c\rfloor_n \lfloor d\rfloor_n \over \lfloor e\rfloor_n
\lfloor f\rfloor_n \lfloor g\rfloor_n \lfloor n\rfloor !} \ \rho^n ,
\]
\beq
\lfloor a\rfloor_n := \lfloor a\rfloor \lfloor a+1\rfloor \cdots
\lfloor a+n-1\rfloor ,
\quad \lfloor a\rfloor_0 :=1
\label{2.21}
\eeq
In the present context, we have
$$
a=m-J,\, b=m_2-J', \, c=-J-m_1, \, d=-n'=-(J'+m'), \,
e=-2x-N-1,
$$
\beq
f=m_2-m'+1, \, g=1+a+b+c+d-e-f \, .
\label{2.22'}
\eeq
To make contact with orthogonal polynomials, we bring the $_4 F_3$ into a
different form by means of the q-version\footnote{it can be proven \cite{JFR}
exactly along the same lines using the method explained,
for instance in ref.\cite{A}  of  the classical formula\cite{S}}
of a well-known transformation formula for balanced $_4 F_3$ functions:
\beqa
_4 F_3\left (^{a,}_{e,} \,
^{b,}_{f,} \, ^{c,}_
{1+a+b+c+d-e-f} \, ^{-n'} \ ;q,1\right )=
{\lfloor f-c\rfloor_{n'} \lfloor f+e-a-b\rfloor_{n'}
\over \lfloor f\rfloor_{n'}\lfloor f+e-a-b-c\rfloor_{n'}}
\times \nnn
_4 F_3\left (^{e-a,}_{e,} \,\,
^{e-b,}_{f+e-a-b,} \,\, ^{c,}_
{1+c-f-n'} \,\, ^{-n'} \ ;q,1 \right )
\label{2.22}
\eeqa
where $n'$ is a non-negative integer. The $_4 F_3$ on the
RHS of  Eq.\ref{2.22} can now be identified with an Askey-Wilson (or
Racah) polynomial\cite{AW}:
$$
p_{n'} (\mu (z);\alpha,\beta,\gamma,\delta;q) =
\ _4 F_3\left (^{e-a,}_{e,} \,\,
^{e-b,}_{f+e-a-b,} \,\, ^{c,}_
{1+c-f-n'} \,\, ^{-n'} \ ;q,1 \right )
 $$
where  \hfill
$$
\mu (z)= q^{-2z} + q^{2(z+c+e-b)} , \, z=-c=J+m_1,
$$
and \hfill
\beq
\alpha =e-1, \, \beta =d-a, \, \gamma =c+d-f, \, \delta =e+f-b-d-1 \, .
\label{2.23}
\eeq
$n'=J'+m'$ is the degree in the variable $\mu (z) $. Note that
$p_{n'}(\mu (z))$ really depends on $z$ only via $\mu (z)$; the other
coefficients are functions only of $J,J'$ and $M$.
The Askey-Wilson polynomials satisfy orthogonality relations of the form
\cite{AW}:
\beq
\sum_{z=0}^{N} p_n(\mu (z))\ p_m(\mu (z))\ w(z)=0
\quad  \hbox{for}\> m \ne n. \label{2.27}\eeq
Here $N=J+J'+M$. The weight function $w(z)$ is defined as
\[
w(z;\alpha,\beta,\gamma,\delta;q)=
\]
\beq
{\lfloor 2z+1+\gamma+\delta\rfloor
\lfloor \gamma+\delta+1\rfloor_z
\lfloor \alpha+1\rfloor_z \lfloor \beta+\delta+1\rfloor_z \lfloor \gamma+1
\rfloor_z
\over \lfloor \gamma+\delta+1\rfloor \lfloor z\rfloor ! \lfloor \gamma+
\delta-\alpha+1\rfloor_z
\lfloor \gamma-\beta+1\rfloor_z \lfloor \delta+1\rfloor_z},
\label{2.28}
\eeq
where $\alpha,\beta,\gamma,\delta$ are the same coefficients as in
the definition of $p_{n'}$.

Thus for any fixed $J,J',m,m'$ we can represent the braiding matrix
$R(J,J';\varpi)_{m_{\phantom{2}} m'}^{m_2 m_1} $ defined
by Eqs.\ref{1.3} and \ref{2.17} -- \ref{2.23}
by the Askey-Wilson polynomial $p_{n'} (\mu (z))$,
up to a prefactor which is independent of $z$. Inserting
into Eq.\ref{2.16},
and putting $y$ equal to one of the zeros of the LHS,
$$
y=-q^{\varpi +2J'-2J+2j_0 -1}, \qquad j_0 =1\cdots J+m
$$
resp. \hfill
\beq
y=-q^{-(\varpi -2J'+2m+2l_0 -1)}, \qquad l_0 =1\cdots J'+m'
\label{2.25}
\eeq
we arrive at the following homogeneous relations:
$$
\noindent
0=\sum_{z=0}^N \lfloor 2x+j_0 +1\rfloor_{N-z}
\lfloor J'-J-M+j_0\rfloor_{z}
{(-1)^z\lfloor -2J\rfloor_{z} \over
\lfloor -2x-2M\rfloor_{z}} p_{n'}(\mu (z)) w(z),
$$
resp. \hfill
$$
0=\sum_{z=0}^N  \lfloor J'-m-M-2x-l_0\rfloor_z
\lfloor J-m-l_0+1\rfloor_{N-z}
{(-1)^z \lfloor -2J\rfloor_{z} \over
\lfloor -2x-2M\rfloor_{z}} p_{n'}(\mu (z)) w(z),
$$
with \hfill
$$
j_0 =1 \cdots J+m,
\qquad
l_0 =1 \cdots J'+m',
\qquad
N=J+J'+M,
\qquad
n'=J'+m'
$$
\beq \label{2.26} \eeq
Next, one proves by induction in $n'=J'+m'$ ($J,J'$ and $N$ fixed)
that the conditions Eqs.\ref{2.26} are equivalent to the relations
Eqs.\ref{2.27}.  Considering the first relation in Eq.\ref{2.26},
we have to show that the q-products  in front of $p_{n'}(\mu (z))w(z)$
are given by a linear
combination of $p_k(\mu (z))$ with $k \ne n'$. Indeed we shall prove
that
$$
\lfloor 2x+j_0'-n'+1\rfloor_{N-z} \lfloor -J-M-m'+j_0' \rfloor_z
{(-1)^z \lfloor -2J\rfloor_z  \over \lfloor -2x-2M\rfloor_z}= \nnn
\sum_{k=n'+1}^N a_{j_0' k} p_k(\mu (z)), \nnn
$$
with \hfill
\beq
j_0'= j_0 +n' = n'+1 \cdots N
\label{2.29}
\eeq

We start at $n'=N-1$ and go downward. For $n'=N-1$, where $j_0'=N$
is the only allowed value, the q-product on the left must be proportional
to $p_N(\mu (z))$. The latter is given by Eq.\ref{2.23}, with the $_4 F_3$
collapsing into a $_3F_2$ since two of the coefficients are equal:
\beq
p_N(\mu (z))=
\ _3 F_2\left (^{-(J'-J+M+2x+1),}_{-(N+2x+1),} \,\,
^{-z,}_{-(2M+2x),} \,\, ^{-(2J+2M+2x+1-z)} \ ;
q,1 \right )
\label{2.30}
\eeq
By means of a well-known relation for the $_3 F_2$ functions\cite{S},
this can be further reduced to
a single product of q-numbers,
so that
\beq
p_N(\mu (z))=\hbox{const}\cdot
\lfloor 2x+2\rfloor_{N-z}\lfloor J'-J-M+1\rfloor_z
{(-1)^z \lfloor  -2J\rfloor_z  \over \lfloor -2x-2M\rfloor_z }
\label{2.31}
\eeq
where the constant is independent of $z$. In view of Eq.\ref{2.29},
this proves the induction start. Consider now the induction step
$m' \rightarrow m'-1, \ m \rightarrow m+1$. This amounts to changing
$j_0' \rightarrow j_0' +1$ in Eq.\ref{2.29}. The case
$j_0' \le N-1$ is covered by the induction
hypothesis. When $j_0'=N$, we have to show that
$$
\lfloor 2x+2N+1-n'-z\rfloor \lfloor J'-m'+z\rfloor
\lfloor 2x+N-n'+1\rfloor_{N-z} \times
$$
\beq
\lfloor -J-M-m'+N\rfloor_z
{(-1)^z  \lfloor -2J\rfloor_z  \over \lfloor -2x-2M\rfloor_z}=
\sum_{k=n'}^N a_k p_k(\mu (z))
\label{2.32}
\eeq
for suitable $a_k$. Using the definition Eq.\ref{2.23} of $\mu (z)$,
it is easy to check that the first two factors in Eq.\ref{2.32}
can be written in the form $c_1 +c_2 \mu (z)$, whereas the others
are known by virtue of the induction hypothesis
to be a linear combination of $p_{n'+1}\cdots p_N$.
But on the other hand the
Askey-Wilson polynomials fulfill a 3-term recurrence
relation of the form\cite{AW}
\beq
\mu (z)p_n(\mu (z))= A_n p_{n+1}(\mu (z)) +B_n p_n(\mu (z)) +
C_n p_{n-1}(\mu (z))
\label{2.33}
\eeq
Since $A_N=0$, the induction step is complete. The proof of the
second relation in Eq.\ref{2.26} is completely analogous.
So we know now
that our conjecture for $R$ is correct up to normalization. The
latter can be easily checked by putting $y=-q^{\varpi +2M-1}$ in
Eq.\ref{2.16}. Then the products on the RHS vanish except when $J+m_1=0$.
Inserting for $R$ as before and using $p_n(\mu (0)) \equiv 1\  \forall\  n$,
it is then a
straightforward exercise to check that Eq.\ref{2.16}
is also fulfilled in this
inhomogeneous case.
Thus we have established that the braiding of the
$V_m^{(J)}$ resp. $U_m^{(J)}$ operators with general $J$, and $J+m$ a
non-negative integer, is given by the extension  Eq.\ref{2.20} or \ref{2.23}
of the R-matrix of \cite{CGR}.
In particular, of course, we reproduce the result for the half-integer case,
without having taken recourse to the null-vector decoupling equation.

As an important consequence of this fact, one obtains that the construction
presented in \cite{G5} for the Liouville exponentials
Eq.\ref{1.1} with positive
half-integer $J$ as local operators possesses an immediate continuation to
arbitrary $J$:
\begin{equation}
e^{\textstyle -J\alpha_-\Phi(\sigma, \tau )}=
\>\sum _{n=0}^{\infty} \,
{B_{n-J}^{(J)}(\varpi)\over
I_{n-J}^{(J)}(\varpi)\bar I_{n-J}^{(J)}(\bar \varpi)}\>
U_{n-J}^{(J)}(x_+)\,
{\overline U_{n-J}^{(J)}}(x_-),
\label{2.34}
\end{equation}
where now one does not stop, since starting from $m=-J$, one does not
reach $m=J$ after integer  steps. One may see that there is no problem
with extending the expression for $B_{n-J}^{(J)}(\varpi)$ to non-integer $J$,
and this defines general exponentials of the Liouville field. Locality
is guaranteed, just as in the half-integer case, by the  fact that
the $R$-matrix for the right-moving chiral fields is essentially the
inverse  of the $R$-matrix for the left movers, as can easily be shown
also for our extension to general $J$.
Details will be given elsewhere.
Finally, we remark that our results can be rephrased in terms of the
fields $\xi _M^{(J)}$ defined in ref.\cite{G1} which form
representations of the
quantum group, and which differ from the $U_m^{(J)}$ by a linear
transformation:
\beq
\xi_M^{(J)} = \sum_m   | \widetilde {J,\varpi })_M^m U_m^{(J)}
\label{2.35}
\eeq
with suitable coefficients
$| \widetilde {J,\varpi })_M^m $. Again, the basis
transformation $| \widetilde {J,\varpi })_M^m $
can be extended straightforwardly
to general $J$. Consequently, also the formula given in  \cite{G5}
for the Liouville exponentials in terms of the quantum group basis
possesses an immediate extension.



\begin{thebibliography}{**}
\bibitem{B} O. Babelon,
\pl   B215,  (1988) , 523.

\bibitem{G1} J.-L. Gervais,  \cmp 130, 257, (1990) .

\bibitem{G2} J.-L. Gervais, \pl B243, 85, (1990) .

\bibitem{CG2} E. Cremmer, J.-L. Gervais,
 \cmp 144, 279, (1992) .

\bibitem{G4} J.-L. Gervais, \ijmp 6, 2805, (1991)
\bibitem{G3} J.-L. Gervais, \cmp 138, 301, (1991) .

\bibitem{G5} J.-L. Gervais, ``Quantum group derivation
of 2D gravity-matter coupling'' Invited talk at
the Stony Brook meeting {\sl String and Symmetry 1991},
LPTENS preprint 91/22,\np B391, 287, (1993).

\bibitem{CGR} E. Cremmer, J.-L. Gervais,
J.-F. Roussel, ``The quantum group structure of
2D gravity and minimal models II:
The genus-zero chiral bootstrap'' LPTENS preprint
93/02, hep-th9302035.

\bibitem{FFK} G. Felder, J. Fr\"ohlich, G. Keller,
\cmp 124, 647, (1989)

\bibitem{JFR} J.F. Roussel, private communication.

\bibitem{GR1} J.-L. Ger\-vais, B. Ros\-tand,
 \np B346, 473, (1990) .

\bibitem{GR2} J.-L. Ger\-vais, B. Ros\-tand,
 \cmp 143, 175, (1991) .

\bibitem{BiG} A. Bilal, J.-L. Ger\-vais,
``Conformal theories with nonlinearly extended Virasoro symmetries
and Lie-algebra classification'' Proceedings of the meeting
{\sl Infinite Dimensional Lie Algebras and Lie Groups}, Marseille
CIRM-Luminy July 1988.

\bibitem{OW} H. Otto, G. Weigt,
\pl B159, 341, (1985) ; { Z. Phys. } {\bf C31},
219, (1986).


\bibitem{LS} D. Luest, J. Schnittger,
\ijmp A6, 3625, (1991)

\bibitem{GN3} J.-L. Gervais, A. Neveu,  \np B224, 329,
(1983).

\bibitem{DF}  Vl. Dotsenko, V. Fateev,
\np B251, 691, (1985) .

\bibitem{KR} A. Kirilov, N. Reshetikhin, {\sl Infinite Dimensional
Lie Algebras and Groups, Advanced Study in
Mathematical Physics} {\bf vol. 7},
 Proceedings of the 1988 Marseille
Conference, V. Kac editor, p. 285, World scientific.

\bibitem{HHM} Bo-yu Hou, Bo-yuan Hou,
Zhong-qi Ma, {\sl Comm. Theor. Phys.}, {\bf 13},  181,
(1990); {\sl Comm. Theor. Phys.}, {\bf 13},  341 (1990).

\bibitem{A}  G. Andrews ``{\sl $q$-series:
their development and application
in analysis, number theory, combinatorics,
physics, and computer algebra}'',
Conference board of the Mathematical Sciences,
Regional Conference in
Mathematics, \# 66, A.M.S. ed.

\bibitem{S} L.C. Slater,
``{\sl Generalized hypergeometric functions}'' Cambridge
 University presss 1966.

\bibitem{AW}R. Askey, J. Wilson,
``{\sl A set of orthogonal polynomials that
generalize the Racah coefficients or 6-j symbols}'' SIAM J. Math. Anal.
 10 (1979) 1008.



\end{thebibliography}
\end{document}